# Frequency Dependence of Superconducting Cavity Q and Magnetic Breakdown Field

**Mario Rabinowitz**
*StanfordLinear Accelerator Center*
*Stanford University, Stanford, California* 94305

Inquiries to: *Armor Research*
*715 Lakemead Way, Redwood City, CA 94062*
Mario715@earthlink.net

**Abstract**

A theoretical explanation is given to account for the unexpected observation that L- and Sband Nb superconducting cavities were found to have lower Q and lower magnetic breakdown field than those of the higher X-band frequencies. Both effects can be related to the trapping of magnetic flux in the cavity walls. The frequency dependence arises from the frequency dependence of the resistivity of oscillating fluxoids. Calculations based on this model are in agreement with experimental observations.

According to established theories [1, 2] of rf superconductivity, one would expect the superconducting surface resistance $R_s$ to be approximately proportional to the square of the cavity angular frequency $\omega$. Hence, in going to the lower frequency L- and S-band Nb cavities, it was expected that the Q's would be as high or higher than with comparably processed X-band cavities. However, as we shall see, this may not be the case when the surface resistance is dominated by trapped flux at the operating temperature. A discussion of the trapping of flux due to an incomplete Meissner-Ochsenfeld effect, and the related power dissipation in an rf superconductor, is given by Rabinowitz. [3]

Assuming that the only nonsuperconducting loss is due to trapped flux, the total average power loss for a magnetic field $H_p \cos\omega t$ at the cavity surface is

$$P = \tfrac{1}{2} \int R_n H_p^2 \, dA_n + \tfrac{1}{2} \int R_s H_p^2 \, dA_s, \tag{1}$$

giving an effective surface resistance for the cavity

$$R = R_n (A_n/A_t) + R_s [1 - (A_n/A_t)], \tag{2}$$

where $R_n$ is the surface resistance of the fluxoids, $A_n$ is the normal area, $R_s$ is the superconducting surface resitance, $A_s$ is the superconducting area, and $A_t$ is the total cavity area.

$$Q = \frac{\frac{1}{2}\int_0^V \mu_0 H_p^2 \, dV}{P/\omega} = \frac{GV\omega}{RA_t} = \frac{GV\omega}{R_n A_n + R_s(A_t - A_n)}, \quad (3)$$

where V is the cavity volume and G is a constant related to cavity geometry. The trapped flux is proportional to the total flux intercepted by the cavity, so that

$$A_n \doteq d(B/B_0)A, \quad (4)$$

where $B$ is the ambient magnetic flux density $H_0 = B_0/\mu$ is the corresponding critical field for type I or II, $A$ is the cross-sectional area of the cavity normal to the flux, and $d$ is a proportionality constant. Therefore,

$$Q = \frac{GV\omega}{R_n(dBA/B_0) + R_s(A_t - dBA/B_0)}. \quad (5)$$

If the power loss is dominated by the trapped flux

$$R_n(dBA/B_0) \gg R_s(A_t - dBA/B_0), \quad (6)$$

and

$$Q \doteq GV\omega B_0/R_n dBA, \quad (7)$$

$$R_n = \rho/2\lambda, \quad (8)$$

where $\lambda$ is the penetration depth and $\rho$ is the effective resistivity of an oscillating fluxoid as derived by Rabinowitz [3], we have

$$\rho = \left(\frac{\omega^2 \phi^4 HH_0 \mu^2}{\rho_n^2(\omega^2 M - p)^2 + \omega^2 \phi^2 H_0^2 \mu^2}\right)\rho_n. \quad (9)$$

$\varphi$ is the flux trapped in a fluxoid, $H$ is the magnetic field in the fluxoid of permeability $\mu$, $H_0$ is the appropriate critical field, $\rho_n$ is the normal state resistivity, $M$ is the fluxoid mass/length, and p is the pinning constant/length. Since the implications of Eq. *(9)* may not be transparent, let us consider some limiting cases.

As previously pointed out [3], when the viscous damping force is negligible,

$$\rho = [\omega^2 \phi^2 HH_0 \mu^2/(\omega^2 M - p)^2](1/\rho_n). \quad (10)$$

When the viscous damping force dominates, then

$$\rho = (H/H_0)\rho_n. \tag{11}$$

Therefore, if the power loss due to trapped flux dominates and a cavity is being operated under the conditions of Eq. (10), more lattice defects within $2\lambda$, consistent with superconducting requirements, may be desirable to reduce the power loss, since this yields a higher $\rho_n$ and $\rho \propto 1/\rho_n$ here. When a cavity is being operated under the conditions of Eq. (11), then higher material purity would be desirable.

Let us consider the effect on $Q$, when cavities are operated in the two regions given by Eq. (10) and in the third region given by Eq. (11). When $\omega^2 M \gg p$, Eqs. (10) and (8) yield

$$R_n = (\phi^2 H H_0 \mu^2 / 2\lambda \rho_n M^2)\omega^{-2}. \tag{12}$$

When $p \gg \omega^2 M$,

$$R_n = (\phi^2 H H_0 \mu^2 / 2\lambda p^2 \rho_n)\omega^2. \tag{13}$$

Combining Eqs. (12) and (7),

$$Q = (2GVB_0\lambda\rho_n M^2 / dBA\phi^2 HH_0 \mu^2)\omega^3. \tag{14}$$

If we were to compare two geometrically similar cavities of different frequency in the same mode and field orientation, assume that they have the same $\rho$ preparation and processing history, neglect any difference in their ability to exclude flux, and any differences in the topography of the trapped flux, then Eq. (14) gives

$$Q \propto \omega^2 B^{-1}, \tag{15}$$

since $V \propto \omega^{-3}$ and $A \propto \omega^{-2}$ for geometrically similar cavities.

Therefore, if one cavity is operated at $8.6$ GHz (X band), the cavity operated at 1.3 GHz (L band) will have its $Q$ lower by a factor of 44 in the same field. The highest reported $Q > 5 \times 10^{11}$ for a 10.5-GHz Nb cavity was measured at SLAC. [4] Turneaure and Viet' reported $Q > 10^{11}$ at $8.6$ GHz. Equation (15) would then predict $Q > 2 \times 10^9$ at 1.3 GHz, all the other factors being similar. This is in good agreement with the HEPL results of Turneaure [6] at 1.3 GHz, in which $Q$ ranged between $2 \times 10^9$ and $4 \times 10^{10}$. All the factors are not necessarily equal or similar; in particular, the ambient magnetic flux density $B$ has been significantly different. Typically, for X-band measurements in shielded degaussed Dewars, $B \sim 10^{-5}$ to $10^{-4}$ G. In the HEPL L-band accelerating structure [6], $B \sim 10^{-3}$ G. In addition to the ambient magnetic field, thermoelectric currents generated during cooldown can also contribute to the trapped flux, as pointed out by Pierce. [7] Niobium S-band measurements from $2.2$ to $3.7$ GHz at SLAC [8] and at other laboratories [9,10] have yielded a range of $Q$'s from $10^9$ to $10^{10}$, also consistent with this theory.

Combining Eqs. (13) and (7), we get

$$Q \doteq (2GVB_0\lambda\rho_n p^2/dBA\phi^2 HH_0\mu^2)\omega^{-1} \qquad (16)$$

Making the same kind of comparison as before, Eq. (14) gives

$$Q \propto \omega^{-2} B^{-1}. \qquad (17)$$

This has the same dependence on $\omega$ as expected from the superconducting surface resistance, and from stationary nonmagnetic normal regions. [11] It appears to be quite advantageous to operate in this region of negligible viscous damping, and dominant pinning, if possible.

Now to consider the region where the viscous damping force dominates, then Eqs. (8) and (11) yield

$$R_n = H\rho_n/2\lambda H_0. \qquad (18)$$

Combining Eqs. (7) and (18), we get

$$Q \doteq (2GVB_0\lambda H_0/dBAH\rho_n)\omega. \qquad (19)$$

Making the same kind of comparison again, Eq. (19) gives

$$Q \propto B^{-1}. \qquad (20)$$

This region has no frequency dependence.

The most general relationship comes from combining Eqs. (7)-(9):

$$Q \doteq \frac{2GV\omega B_0\lambda}{dBA\rho_n}\left(\frac{\rho_n^2(\omega^2 M - p)^2 + \omega^2\phi^2 H_0^2\mu^2}{\omega^2\phi^2 HH_0\mu^2}\right). \qquad (21)$$

Now that we have considered the frequency dependence of $Q$ when it is dominated by the trapped flux power loss, let us also consider the magnetic breakdown field $H_p'$ in this case. As derived by Rabinowitz [3], when breakdown is dominated by fluxoid power loss,

$$H_p' = \left[\frac{-k_{1n}}{2B}\frac{T_c^2}{H_0} + \left(\left(\frac{k_{1n}}{2B}\frac{T_c^2}{H_0}\right)^2 - 2NRF_1^2 b\right.\right.$$

$$\left.\left.\times\left\{\tfrac{1}{2}k_{1n}\left[T_b^2 - \frac{T_c^2}{B}\left(1 - \frac{H_a}{H_0}\right)\right]\right\}\right)^{1/2}\right](NRF_1^2 b)^{-1}. \qquad (22)$$

for the case of a fluxoid perpendicular to the surface. An equation of the same form is derived for a parallel fluxoid [2,8]. For the present purposes, in which we are primarily concerned with the frequency dependence of $H_p'$, let us substitute the combination of

Eqs. (8) and (10) into Eq. *(22)*, representing most of the nonfrequency dependent terms and factors by $k_i$:

$$H'_p = \left[ -k_1 + \left( k_1^2 + \frac{k_2 \omega^2}{(\omega^2 M - p)^2 + k_3 \omega^2} \right)^{1/2} \right]$$
$$\times \left( \frac{k_4 \omega^2}{(\omega^2 M - p)^2 + k_3 \omega^2} \right)^{-1}. \qquad (23)$$

Let us consider the three regions of $R_n$ again. If $(\omega^2 M - p)^2 \gg k_3 \omega^2$ and $\omega^2 M \gg p$, then

$$H'_p = \left[ -k_1 + (k_1^2 + k_2 M^{-2} \omega^{-2})^{1/2} \right] k_4^{-1} M^2 \omega^2. \qquad (24)$$

This is for the case of negligible viscous damping and negligible pinning. When $k_3 M^{-2} \omega^{-2} \gg k_1^2$, then

$$H'_p \propto \omega. \qquad (25)$$

Therefore, in going from 8.6 to 1.3 GHz, the magnetic breakdown field could be reduced by a factor of 6.6. Turneaure and Viet [5] reported a breakdown field of 1080 Oe at 8.6 GHz. So, if the conditions governing Eq. *(25)* were met, $H_p'$ would be ~160 Oe for a 1.3-GHz cavity and ~300 - 400 Oe for S-band. Values of ~*300* Oe have been obtained at 1.*3* GHZ 6 and ~ 200 - 400 Oe at S band. [8 - 10] Even aside from the question of whether the conditions of Eq. *(25)* apply, it must be borne in mind that breakdown is dominated by the fluxoid in the most vulnerable position. [3,12] It is not too likely that two cavities will have the dominant fluxoid in the same position. Nevertheless, it is significant that Eq. *(25)* gives $H_p'$ within a factor of *2* of the experimental value.

When the damping is negligible and the pinning is dominant, $(\omega^2 M - p)^2 \gg k_3 \omega^2$ and $p \gg \omega^2 M$, then Eq. *(23)* becomes

$$H'_p = \left[ -k_1 + (k_1^2 + k_2 p^{-2} \omega^2)^{1/2} \right] p^2 k_4^{-1} \omega^{-2}. \qquad (26)$$

This would be a nice region to work in, if possible, both for high $H_p'$ and high Q.

When the viscous damping force dominates $k_3 \omega^2 \gg (\omega^2 M - p)^2$ and $H_p'$ has no frequency dependence.

Besides trapped flux, other factors may well enter in to complicate the situation. The larger surface area of the lower-frequency cavities increases the occurrence probability of protrusions on the cavity surface which can enhance the electric and magnetic fields locally. The larger cavities having the same wall thickness as the smaller X-band cavities are subject to more mechanical strain during cooldown after annealing, and simply during operation. This can increase the probability of protrusion growth. The growth of large crystallites spanning the wall thickness with grain boundaries largely parallel to the direction of heat flow increases the thermal conductivity of the polycrystalline material to values comparable to that of single crystals, which increases the high-power Q and magnetic breakdown field. [3, 12] Yet this may be offset, because the super-conducting -state thermal conductivity is reduced by strain. The presence of a second (smaller) energy gap in the (110) and (111) planes in niobium can yield more power loss for transport currents in these directions. Rolled Nb (usually used for large cavities) can preferentially have these planes parallel to the conducting surface, and high -temperature annealing may not totally erase this preferential lattice orientation. Hence, differences in the processing of small cavities (usually machined out of solid Nb) and large cavities may be a factor. Nevertheless, it would appear that calculations based on the model of trapped flux dominating the frequency dependence of cavity Q and magnetic breakdown field are in good accord with experimental observations.

*Work supported by the U.S. Atomic Energy Commission.